\documentstyle[aps,prb,epsf]{revtex}

\begin{document}
\draft
\preprint{ETH-TH/97-12}

\title
{
\LARGE \bf
Decay of Metastable States:
Sharp Transition from Quantum to 
Classical Behavior}
\author{{\bf D. A. Gorokhov  and  G. Blatter}
\\{\it Theoretische Physik, ETH-H\"onggerberg,
CH-8093 Z\"urich, Switzerland}\\
{\it e-mail: gorokhov@itp.phys.ethz.ch} }
\maketitle
\begin{abstract}
The decay rate of metastable states is determined at high temperatures by
thermal activation, whereas at temperatures close to zero
quantum tunneling is relevant. At some temperature
$T_{c}$ the transition from
classical to quantum-dominated decay
occurs. The transition can be
first-order like, with a discontinuous first derivative
of the Euclidean action,
or smooth with only a second derivative  developing a jump.
In the former case the crossover temperature 
$T_{c}$ cannot be calculated
perturbatively and must be found as the intersection point of the
Euclidean actions calculated at low and high temperatures.
In this paper we present a 
sufficient criterion for a first-order transition
in tunneling problems and apply it to the problem
of the tunneling of strings. It is shown that the problem of the 
depinning of a massive string from a linear defect in the presence of an
arbitrarily strong dissipation exhibits a first-order transition.
\end{abstract}
\pacs{PACS numbers: 64.60.My, 74.50.+r, 74.60.Ge} 

\section{Introduction}
Investigations of the decay rate of metastable states have a long history,
 going back to Kramers\cite{Kramers}
who calculated the lifetime of a classical 
trapped particle separated from the
true equilibrium state by a high potential barrier.
Since then different decay phenomena have been investigated.
The motion of dislocations across
the Peierls barrier,\cite{Petukhov,Ivlev}
the decay of the current in a Josephson loop,\cite{Larkin} and
the creep of vortices in
superconductors\cite{Blatter}
are typical and well-known examples of metastability.
At sufficiently high temperatures a metastable system decays due to thermal activation and the decay rate $\Gamma$  obeys the Arrhenius law
$\Gamma\sim\exp{(-U/T)},$ whereas at temperatures close to absolute zero  quantum tunneling is relevant and $\Gamma\sim\exp{(-S/\hbar)},$
with $S$ the action at zero temperature. Below we shall consider
systems for which the semiclassical description is applicable. In this case
$U/T$ and $S/\hbar$ are large; otherwise, the system would not be
truly metastable.

Let ${\bf u}(t,{\bf r})$ denote the coordinates of the system
under consideration
depending in general on imaginary time $t$ and spatial variables
${\bf r}.$ Within the semiclassical approximation the decay
rate at a temperature $T$ is determined by the contribution
 of the trajectories close to the one
extremizing the Euclidean action and satisfying the periodicity
condition ${\bf u}(0,{\bf r})={\bf u}(\hbar /T, {\bf r}).$
In the high-temperature regime the function extremizing the action is
time independent and,
consequently, the activation barrier in the Arrhenius law $U$ does not depend on
the dynamic properties of the system, meanwhile the bounce solution
at low temperatures
is time-dependent and the dynamics enters.
At some temperature $T_{c}$ the transition from the time-independent
to the time-dependent
solution occurs. In some cases the bounce solution just below the crossover point can be written in the form

\begin{equation}
{\bf u}(t,{\bf r})={\bf u}_{th}({\bf r})+\delta{\bf u}({\bf r})
\cos\left (\frac{2\pi T_{c}t}{\hbar}\right ),
\label{40}
\end{equation}
where the function $\delta {{\bf u}({\bf r})}$ is considered to be small.
Such an assumption leads to an action $S_{\rm Eucl}(T)$ with a
continuous
first derivative
in the crossover point (see Fig.~1).

The second derivative of $S$ is discontinuous at the point $T_{c}.$
The second derivative has a jump at $T=T_{c}.$ Following
Ref.\cite{Ovchinnikov}
we shall call this situation a ``second-order transition
in the crossover point."
In this case the bounce solution
with the minimal Euclidean action
at low temperatures can be obtained by a continuous deformation
of the thermal solution. However, in general such a deformation
is not possible: The trajectory corresponding to the minimal action may
jump at a certain temperature. In this case we deal with a first-order
transition: it can be shown that the first derivative
of $S_{\rm Eucl}(T)$ has a jump at the transition point (see Fig.~2)
and the expansion
$(\ref{40})$ is not valid.

Strictly speaking, even if we deal with a first-order transition
in a tunneling problem, there is a narrow crossover region
from one solution to another because
contributions originate from
several
saddle points and we need to take into account
all of them. However, the better the semiclassical approximation is
applicable, the narrower this region becomes.

The problem 
considered here
is related to the mean-field theory of phase transitions. The Euclidean action
can be identified with the free
energy in the Landau theory with an order parameter
$\delta({\bf r},T)$
defined as 
$\delta ({\bf r},T)=\max_{t} |{\bf u}(t, {\bf r})-{\bf u}_{th}({\bf r})|,$
for example.
At high temperatures
${\bf u}(t, {\bf r})={\bf u}_{th}({\bf r})$
 and
$\delta({{\bf r},T})\equiv 0,$
whereas at $T<T_{c},$ $\delta({\bf r},T)\ne 0.$
The order parameter
changes continuously at the point $T=T_{c}$ in the case of a second-order
transition and discontinuously if a first-order transition takes place.

\centerline{\epsfxsize=4cm \epsfbox{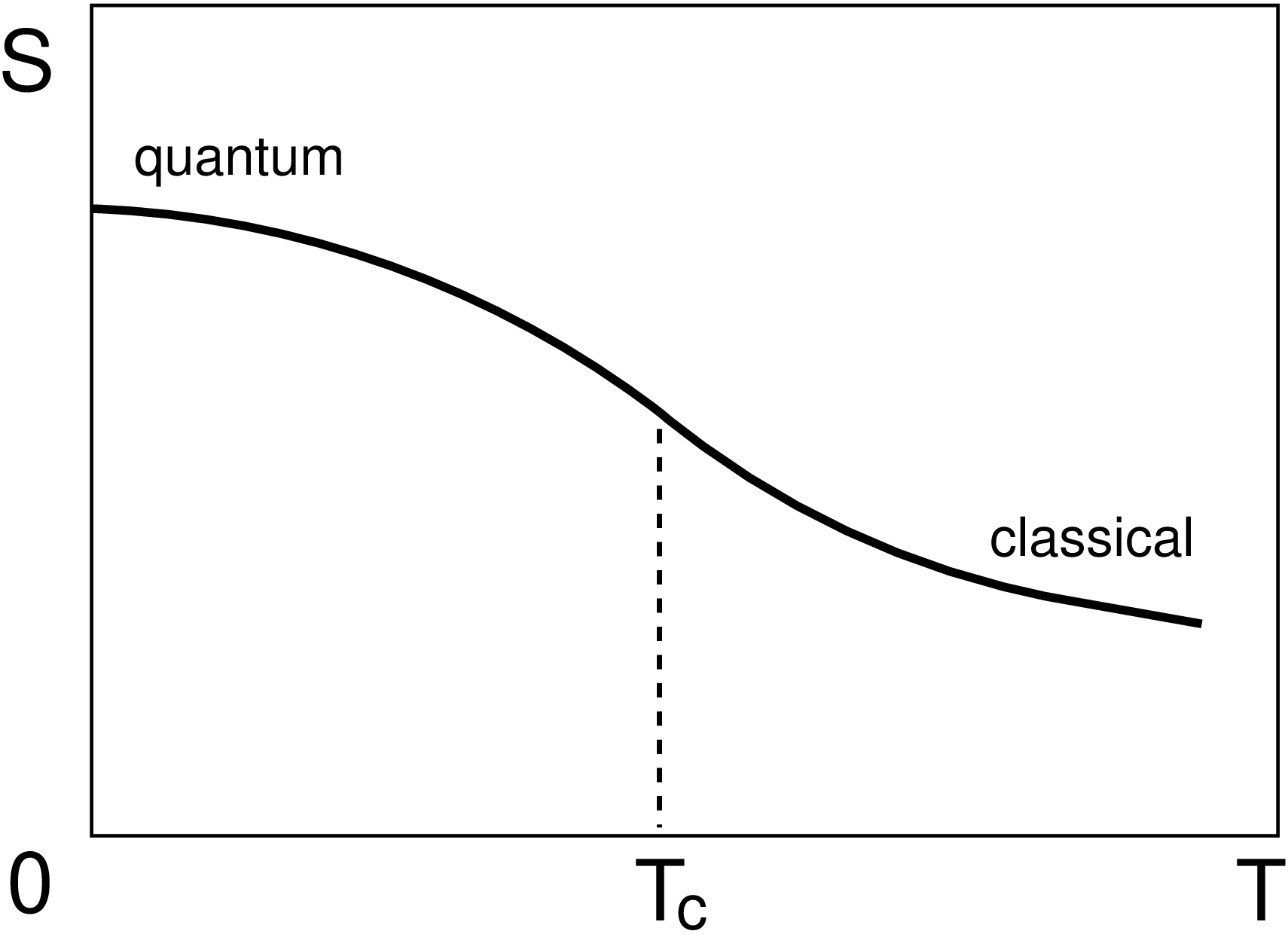}}
{\footnotesize {\bf Fig.1}~Euclidean action $S$ as a function of temperature
$T$  for the case of
a second-order transition.}
\vskip0.3cm
\hskip-0.7cm

\centerline{\epsfxsize=4cm \epsfbox{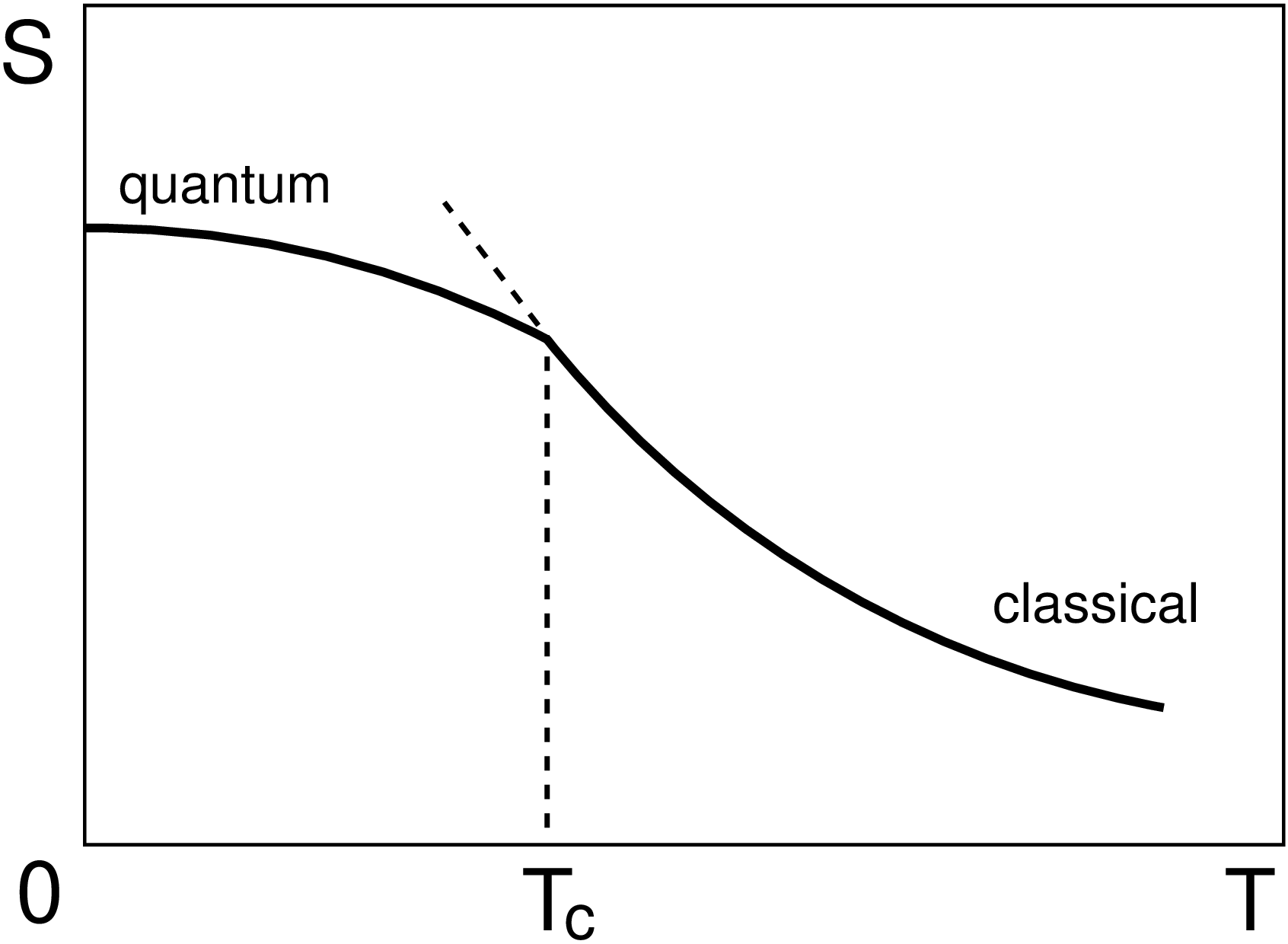}}
{\footnotesize {\bf Fig.2}~Euclidean 
action $S$ as a function of temperature
$T$ for the case of a
first-order transition. The first derivative of $S$ is discontinuous at the
point $T_{c}.$}
\vskip0.3cm
\hskip-0.7cm


In this paper we present a simple criterion for 
the appearance of a first-order transition
from quantum to classical behavior in tunneling problems
and apply it to the problem of the tunneling of strings.

\section
{Oscillations close to the thermal saddle-point solution}
In this section we study the behavior of the imaginary time oscillation period
of the solution of the equation of motion close to the thermal 
saddle-point solution. In section
\ref{general} we describe the general theory for 
one-dimensional
(1D) Hamiltonian systems.
We briefly summarize the 
results of Lifshitz and Kagan\cite{Lifshitz} 
and Chudnovsky\cite{Chudnovsky} for 1D Hamiltonian systems
and provide
an expression for the jump of the derivative of the Euclidean action
(see Eq.~(\ref{jump}))
for the case of a first-order transition.
 In section \ref{nonlinear}
we use a perturbative approach for the calculation of the imaginary
time oscillation period in the vicinity of the thermal saddle-point solution,
and define 
a criterion for a first-order transition.

\subsection{General theory for 1D Hamiltonian systems}
\label{general}
As shown by Chudnovsky\cite{Chudnovsky}, if the 
imaginary time oscillation
period of a massive particle is not a monotonous
function of energy, a first-order transition from
quantum to classical behavior takes place. This statement
remains true for any one-dimensional Hamiltonian system:
Consider a 1D system subject to 
potential which allows tunneling in one direction (Fig.~3).

\centerline{\epsfxsize=7cm \epsfbox{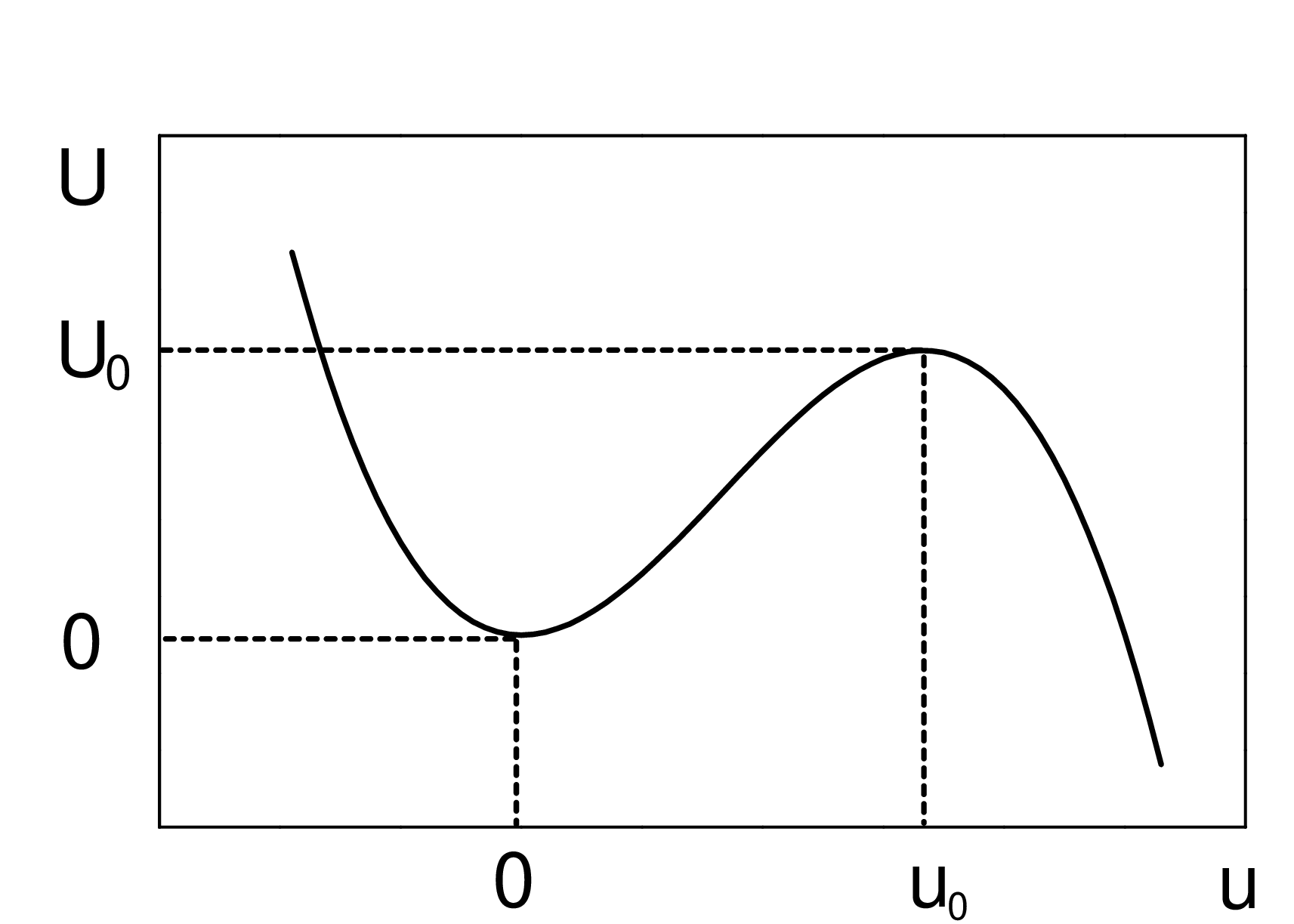}}
{\footnotesize {\bf Fig.3}~1D potential
well with a metastable state. A particle
trapped at $u=0$ can tunnel out to the
right
in order
to reach the true ground state.}
\vskip0.3cm
\hskip-0.7cm
We choose the metastable minimum of the potential as our zero energy.
Let us define the height of the potential barrier as $U_{0}.$
Using the semiclassical approximation we can write the imaginary
parts $\hbar\Gamma_{n}$ 
of the energy $E_{n}$ of the eigenstates in the form\cite{Galitskiy}
\begin{equation}
\hbar\Gamma_{n}=\frac{\hbar \omega ({E_{n}})}{4\pi}
\exp\left(-\frac{2}{\hbar}\int_{a_{n}}^{b_{n}}|p| dx\right)
\equiv
\frac{\hbar \omega(E_{n})}{4\pi}\exp\left(-\frac{S_{n}}{\hbar}\right),
\end{equation}
where $p$ is the generalized momentum,
$E_{n}<U_{0}$ are the energy levels in the
equivalent potential without
tunneling, and $\omega(E_{n})$ are the
corresponding classical oscillation frequencies.
At finite temperature $T$ the decay rate can be easily found by
averaging over the Boltzmann distribution
\begin{equation}
\Gamma=\frac{2}{Z}\sum_{n}\frac{ \omega(E_{n})}{4\pi}
\exp\left (-\frac{E_{n}}{T}-\frac{S_{n}}{\hbar}\right ),
\label{45}
\end{equation}
where
$Z=\sum_{n}\exp\left (-\frac{E_{n}}{T} \right )$
is the partition function of the particle in the well.
With Eq.~$(\ref{45})$ we reproduce the well-known expression
$\Gamma=(2/{\hbar}){\rm Im}F$
for the decay rate at low $T$ (see Ref.~\cite{Tunn}).

Equation $(\ref{45})$ is applicable only at sufficiently low temperatures.
At high temperatures the excitations with energies larger than the barrier
height are relevant in the determination 
of the preexponential factor.
The decay rate then is given by the equation\cite{Kramers,Mel'nikov,Meshkov}
\begin{equation}
\Gamma=
\frac{ \omega (E_{0})}{2\pi}
\exp\left (-\frac{U_{0}}{T}\right ),
\label{high}
\end{equation}
where $\omega(E_{0})$ is the oscillation frequency of the
system near the metastable minimum.
It is necessary to emphasize that the above expressions for the
decay rate are applicable in the
case of intermediate friction $\eta.$ 
E.g., for a particle
of mass $m$ 
the criterion
${\omega T}/{U_{0}} \ll {\eta}/{m}\ll \omega ,$
where ${\omega}^{2} ={-U^{\prime\prime}(u_{0})}/{m}$
with $u_{0}$ the position of the maximum of the metastable
potential
(see Refs.\cite{Kramers,Mel'nikov,Meshkov}),
guarantees, that on the one hand the system in the well is properly
equilibrated, while on the other hand the dynamics is not affected by
the damping.

The sum in Eq.~$(\ref{45})$
can be well approximated by the term where the function
$f_{T}(E)=-{E}/{T}-{S(E)}/{\hbar}$ takes its maximal value and
the problem reduces to the calculation of the maximum
 of $f_{T}(E)$ within the interval
$[0,U_{0}].$
The extremal condition  for $f_{T}(E)$ takes the form
\begin{equation}
\frac{df_{T}}{dE}=-\frac{1}{T}-\frac{1}{\hbar}\frac{d S}{d E}=0.
\label{imp}
\end{equation}

It is well known from standard courses of classical mechanics\cite{Landau}
that
for a periodic problem
the derivative of the action with respect to the energy
is equal to the oscillation time $\tau$ at this energy. In our case
the energy axis has the opposite direction (the action $S(E)$
corresponds to the motion in the inverse potential with energy $E$). Hence,
\begin{equation}
\frac{d S}{d E}=-\frac{\hbar}{T}=-\tau (E),
\label{period}
\end{equation}
 i.e., the decay is dominated by the
Euclidean action with the trajectory periodic in
the time interval ${\hbar}/{T}.$

If the function $\tau (E)$ is a monotonously decreasing function
of energy, Eq.~(\ref{period}) has one solution for
$T<{\hbar}/{\tau_{0}}$ ($\tau_{0}=\tau (U_{0}) $) and no solutions for
$T>{\hbar}/{\tau_{0}}.$ At temperature $T_{c}={\hbar}/{\tau_{0}}$
a second-order transition takes place.

Next, let us suppose that $\tau (E)$ goes through a minimum in the interval
$(0,U_{0})$ (see Fig.~4). 
In this case it can be shown\cite{Chudnovsky}
that there is an energy $E_{1}\in (0,E_{\rm min}),$ and associated
with it a temperature $T_{c}={\hbar}/{\tau (E_{1})},$
where a first-order transition 
takes place.

Let us show that if $E_{1}\ne U_{0},$
the Euclidean action 
$S_{\rm Eucl}(T)$ determining the decay rate $\Gamma$
indeed has a discontinuous derivative
at the point $T=T_{c}$ (see Eq.~(\ref{high})).
At $T>T_{c}$
\begin{equation}
\frac{S_{\rm Eucl}(T)}{\hbar}=
\frac{S_{\rm thermal}(T)}{\hbar}=\frac{U_{0}}{T}.
\end{equation}

\centerline{\epsfxsize=7cm \epsfbox{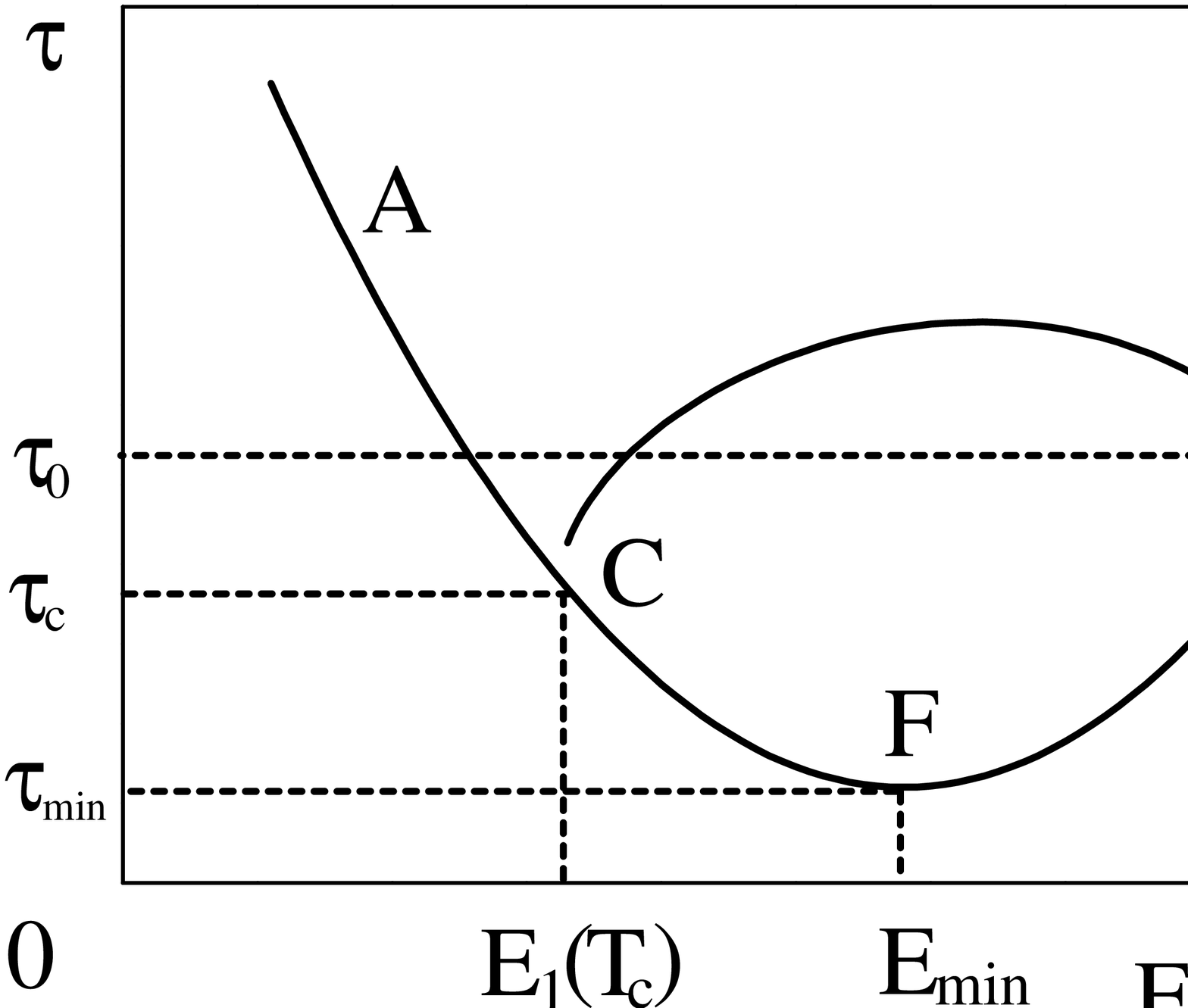}}
{\footnotesize {\bf Fig.4}~{
Non-monotonous
dependence of the oscillation time $\tau$ in the inverse potential
on the energy $E.$ The arrow indicates the jump of the semiclassical trajectory
from the thermal assisted quantum solution to the classical one.}
}
\vskip0.3cm
\hskip-0.7cm
The action just below $T_{c}$ can be written as (see Eq.~(\ref{45}))
\begin{equation}
\frac{S_{\rm Eucl}(T)}{\hbar}=
\frac{S_{\rm quantum}(T)}{\hbar}=\frac{E_{1}(T)}{T}+\frac{S(E_{1})}{\hbar}
\end{equation}
(note that $E_{1}$ is a function of temperature).

The variations of $S_{\rm thermal}$ and $S_{\rm quantum}$ with respect to
the $T$ near the point ${T_{c}}$ are
\begin{equation}
\frac{\delta S_{\rm thermal}}{\hbar}=
-\frac{U_{0}}{{T_{c}}^{2}}\delta T,
\end{equation}
\begin{equation}
\frac{\delta S_{\rm quantum}}{\hbar}=
-\frac{E_{1}(T_{c})}{{T_{c}}^{2}}\delta T
+\bigg (\frac{1}{T_{c}}+\frac{1}{\hbar}\frac{d S}{d E}
\bigg | _{E=E_{1}(T_{c})}\bigg )
\frac {dE_{1}}{dT}\delta T.
 \end{equation}
Taking into account that at the point $E_{1}$
\begin{equation}
\frac {d f_{T}}{d E}\bigg | _{E=E_{1}}=
-\frac{1}{T_{c}}-\frac{1}{\hbar}
\frac{d S}{d  E}\bigg | _{E=E_{1}}=0
\end{equation}
we obtain
\begin{equation}
\frac{\delta S_{\rm quantum}}{\hbar}=-\frac{E_{1}}{{T_{c}}^{2}}\delta T.
\end{equation}
Thus, the jump of the derivative at the point $T_{c}$ is
\begin{equation}
\Delta \frac{dS_{\rm Eucl}}{dT}=
\frac{\hbar \left(E_{1}-U_{0}\right )}{{T_{c}}^{2}}< 0,
\label{jump}
\end{equation}
and the thermal action always decays more rapidly
than the quantum one.
We thus have shown that
if the imaginary time oscillation time
$\tau$  is not a monotonous function of energy $E,$
a first-order transition
takes place. Equation (\ref{jump})
quantifies the strength
of
the first-order transition: if $E_{1}(T_{c})\simeq U_{0},$
the difference in the bounce solutions is small
and the first-order transition is weak. In the opposite with case
$E_{1}(T_{c})\simeq 0$ the saddle-point solution deforms significantly
and we have a strong first-order transition.

For a 1D massive particle the above physics can be easily realized. Consider
a given function $\tau (E)$ for a trapped particle. Then there exist
infinitely many potentials reproducing $\tau (E)$ as their
oscillation times \cite{Landau}.
A simple way to obtain a non-monotonous behavior is
to choose a function $\tau (E)$ with one minimum.
For more complicated nonmonotonous dependences
the system can exhibit several transitions.
In general a system may exhibit several first-order transitions
but not more than one of second-order.

All the results obtained in this section remain true
for an arbitrary 1D metastable Hamiltonian system as Eq.~(\ref{period})
is applicable in this case.

\subsection{Nonlinear perturbation near the thermal saddle-point solution}
\label{nonlinear}
Consider a metastable system whose oscillation period in the imaginary time
$\tau$
is a function of a scalar parameter $a,$ 
$\tau=\tau (a).$ For Hamiltonian systems
it is convenient to choose this parameter to be equal to the energy, i.e.,
$a=E,$ see section~\ref{general}.
However, for dissipative systems one cannot use the energy as it
is no more a conserved quantity. 
In this case we can parametrize the periodic imaginary
time solutions of the equation of motion by the amplitude  $a$ of 
the oscillations:
The amplitude $a$ quantifies the difference between the thermal
 and the weakly time-dependent solutions in the vicinity
of the thermal saddle-point solution
(later we will show how to define such a notion as the ``amplitude''
for any metastable system). 
It can be easily understood that if $\tau (a)$ is
not a monotonous function of the amplitude, the system
exhibits a first-order transition:
Starting from the zero temperature bounce solution
the period $\tau ={\hbar}/{T}$ decreases with increasing temperature.
However we cannot carry this solution beyond the temperature
${\hbar}/{\tau_{min}}$ and thus will encounter a first-order jump
to the thermal solution at some temperature $T_{c}<{\hbar}/{\tau_{min}}$
(in the simplest case of a $\tau$-dependence with one minimum).

The basic idea leading to the criterion
for a sharp transition from quantum to classical behavior
 is the following: Let us investigate
the imaginary time oscillation period 
in the vicinity of the thermal saddle-point, where $a=0.$ 
From the above discussion follows that if
$\tau (a)-\tau (0) < 0,$
the system will exhibit a first-order transition. It turns out that
a general expression for $\tau (a)$ can be obtained
 for a wide class of systems.

At high temperatures the saddle-point
solution is time-independent. Let us slightly perturb this solution and
calculate its oscillation time in the vicinity of the point $a=0.$
We suppose that the imaginary time Lagrangian $L$ of the system
under consideration
can be written in the form
\begin{equation}
L=
T({\bf u},{\bf \dot u},{\bf r})+
V({\bf u},{\bf r}),
\label{ld}
\end{equation}
where the term
$T({\bf u},{\bf \dot u},{\bf r})$
is responsible for the dynamical properties of the system and 
$V({\bf u},{\bf r})$ is the potential energy.
The equation of motion corresponding to the Lagrangian
(\ref{ld}) takes the form
\begin{equation}
{\hat l}{\bf u}=\frac{\delta V}{\delta {\bf u}}.
\label{eq}
\end{equation}
Later we shall suppose that the operator ${\hat l}$
is linear and that ${\delta {\hat l}}/{\delta {\bf r}}=0.$
This is not a strong restriction as the  
dynamical terms traditionally considered 
in the Lagrangian (\ref{ld}) consist of dissipative, massive, and Hall terms.
Dissipation is usually described by the 
Caldeira-Leggett
formalism\cite{Caldeira},  leading
to a linear term in the equation of motion. Massive and Hall terms
also lead to linear equations, i.e., the operator $\hat l$
satisfies the conditions discussed above. For a massive particle
the operator ${\hat l}$ takes the form ${\hat l}=m{\partial^{2}}/{\partial t^{2}}.$

In the high-temperature regime the solution of 
Eq.~(\ref{eq}) ${\bf u}_{th}({\bf r})$ is time-independent.
Let us expand Eq.(\ref{eq}) into a series around this solution.
Substituting
${\bf u}({\bf r},t)={\bf u}_{th}({\bf r})+\delta {\bf u}({\bf r},t)$
into Eq.~(\ref{eq}) and expanding in $\delta{\bf u}$
we obtain
\begin{equation}
{\hat l}\delta {\bf u}=
{\hat h}\delta {\bf u}+
{\hat G}_{2}[{\delta\bf u}]+
{\hat G}_{3}[{\delta\bf u}],
\label{ryad}
\end{equation}
where ${\hat l}={\delta^{2}V}/{\delta {\bf u}^{2}}$ is a linear
operator and
${\hat G}_{2}$ and ${\hat G}_{3}$ are nonlinear operators
satisfying the conditions
\begin{equation}
{\hat G}_{2}[\lambda{\bf y}] ={\lambda}^{2}{\hat G}_{2}[ {\bf y}]
\ \ \ {\rm and}\ \ \
{\hat G}_{3}[\lambda{\bf y}] ={\lambda}^{3}{\hat G}_{3}[{\bf y}],
\label{cond}
\end{equation}
where ${\bf y}$ is an arbitrary vector and $\lambda$ is a constant.
Our goal is to solve Eq.~(\ref{ryad}) 
for $\delta{\bf u}({\bf r},t)$ and find the correction to
the oscillation period away from the thermal saddle point. 
To lowest order in  perturbation theory we separate variables with 
the {\it Ansatz}
\begin{equation}
\delta {\bf u}=a {\bf u}_{0}({\bf r})\cos{\omega_{0}t}
\label{lowest}
\end{equation}
and substituting (\ref{lowest}) into (\ref{ryad}) while neglecting terms of
order higher than $a^{2}$ and higher we obtain
\begin{equation}
{\bf u}_{0}({\bf r})
\left [
{\hat l}\cos{\omega_{0}t}\right ]=
\left [{\hat h}{\bf u}_{0}({\bf r})\right ]
\cos{\omega_{0}t}.
\end{equation}
The function $\cos{\omega_{0}t}$ is an eigenfunction of the operator
${\hat l},$ ${\hat l}\cos{\omega_{0}t}=l(\omega_{0})\cos{\omega_{0}t},$
and we obtain the eigenvalue equation for
${\bf u}_{0}({\bf r})$
\begin{equation}
l(\omega_{0}){\bf u}_{0}({\bf r})=
{\hat h}{\bf u}_{0}({\bf r}).
\label{eig}
\end{equation}

The operator ${\hat h}$ has one negative eigenvalue
due to the unstable direction in phase space. On the other hand,
$l({\omega}_{0})<0,$ i.e., Eq.~(\ref{eig}) has only one solution.
The function ${\bf u}_{0}({\bf r})$ then is the eigenfunction of the operator
 ${\hat h}$
corresponding to its lowest eigenvalue $h_{0}$ and
the frequency $\omega_{0}$ is determined by the condition
$l(\omega_{0})=h_{0}.$ 
Note that $T_{c}={\hbar \omega_{0}}/{2\pi}$ is the transition temperature
in case of a second-order transition.
At this stage it becomes clear how to define
such a notion as an ``amplitude'' for any tunneling problem:
it is the expansion coefficient of the unstable mode.

Next, let us write
\begin{equation}
\delta {\bf u}=a {\bf u}_{0}({\bf r})\cos{\omega t}+
\delta{\bf u}_{2} \left ({\bf r}, t\right ),\ \ \  {\omega}=\omega_{0}+\omega_{2},
\label{sec}
\end{equation}
where ${\omega}_{2}$ is the correction to the frequency ${\omega_{0}}$
and $\delta {\bf u}_{2}\sim a^{2}.$
Substituting (\ref{sec}) into (\ref{ryad}), neglecting terms
of order $a^{3}$ and higher, and using (\ref{cond}) we obtain
the equation for $\delta{\bf u}_{2}$
\begin{equation}
l(\omega)a{\bf u}_{0}({\bf r})\cos{\omega t}+{\hat l}\delta {\bf u_{2}}=
l(\omega_{0})a{\bf u}_{0}({\bf r})\cos{\omega t}+{\hat h}\delta {\bf u}_{2}+
a^{2}
{\hat G}_{2}[{\bf u}_{0}({\bf r})]
{\cos^{2}{\omega t}}.
\label{seceq}
\end{equation}
Rearranging terms, we arrive at
\begin{equation}
\delta{\bf u}_{2}=
{\left ({\hat l}-{\hat h}\right )}^{-1}
\left [
\left (l(\omega_{0})-l(\omega)\right )a{\bf u}_{0}
({\bf r})\cos{(\omega t)}
+\frac{a^{2}}{2}{\hat G}_{2}[{\bf u}_{0}(\bf r)]\left (1+\cos{(2\omega t)}
\right )
\right ].
\end{equation}
Since $({\hat l}-{\hat h}){\bf u}_{0}({\bf r})\cos{\omega t}=0,$
the first term has to vanish and we obtain no shift in the frequency, 
$l(\omega )=l(\omega_{0}),$ i.e., $\omega =\omega_{0},$
$\omega_{2}=0.$
The solution of Eq.(\ref{seceq})
then reads
\begin{equation}
\delta {\bf u}_{2}=
{\bf g}_{1}({\bf r})+{\bf g}_{2}({\bf r})\cos{2\omega_{0}t},
\end{equation}
where 
\begin{equation}
{\bf g}_{1}({\bf r})=
-\frac{1}{2}a^{2}{\hat h}^{-1}{\hat G}_{2}[{\bf u}_{0}({\bf r})],
\end{equation}
\begin{equation}
{\bf g}_{2}({\bf r})=
-\frac{1}{2}a^{2}
{\left ({\hat h}-l(2\omega_{0})\right )}^{-1}
{\hat G}_{2}[{\bf u}_{0}({\bf r})].
\end{equation}

To third order in perturbation theory we make the {\it Ansatz}
\begin{equation}
\delta {\bf u}=a {\bf u}_{0}({\bf r})\cos{\omega t}+
\delta{\bf u}_{2}({\bf r},t)
+\delta{\bf u}_{3}({\bf r},t),
\ \ \  {\omega}=\omega_{0}+\omega_{3}.
\label{third}
\end{equation}
Substituting (\ref{third}) into (\ref{ryad}) and neglecting terms of order
$a^{4}$ and higher we obtain the equation for $\delta {\bf u}_{3},$
\begin{eqnarray}
& \ &   l(\omega)a{\bf u}_{0}({\bf r})\cos{\omega t}+
l(2\omega){\bf g}_{2}({\bf r})\cos{2\omega t}+{\hat l}\delta {\bf u}_{3}
\nonumber\\
& = & l(\omega_{0})a{\bf u}_{0}({\bf r})\cos{\omega t}+
{\hat h}{{\bf g}_{1}}+{\hat h}{\bf g}_{2}\cos{2\omega t}+
{\hat h}{\delta {\bf u}_{3}}\nonumber\\
& + &{\hat G}_{2}
\left [a{\bf u}_{0}({\bf r})\cos{\omega t}+{\bf g}_{1}+
{\bf g}_{2}\cos{2\omega t}\right ]+
{\hat G}_{3}
\left [a{\bf u}_{0}({\bf r})\cos{\omega t}\right ].
\end{eqnarray}

As ${\bf g}_{1}({\bf r}),{\bf g}_{2}({\bf r})\sim a^{2},$
 we can expand the term
${\hat G}_{2}
\left [a{\bf u}_{0}({\bf r})\cos{\omega t}+{\bf g}_{1}+
{\bf g}_{2}\cos{2\omega t}\right ]$
around the function $a{\bf u}_{0}({\bf r})\cos{\omega t}:$
\begin{eqnarray}
& {\hat G}_{2} &
\left [a{\bf u}_{0}({\bf r})\cos{\omega t}+{\bf g}_{1}({\bf r})+
{\bf g}_{2}({\bf r})\cos{2\omega t}\right ]\cong
\nonumber\\
& {\hat G}_{2} &
\left [a{\bf u}_{0}({\bf r})\cos{\omega t}\right ]+
a
\cos{\omega t}
\frac{\delta {\hat G}_{2}}{\delta {\bf u}}
\bigg |_{{\bf u}={\bf u}_{0}({\bf r})}
\left ({\bf g}_{1}({\bf r})
+{\bf g}_{2}({\bf r})\cos{2\omega t}\right ).
\end{eqnarray}
Note that the opeator 
${\delta {\hat G}_{2}}/{\delta {\bf u}}|_{{\bf u}_{0}}$ is a 
linear operator, see Eq.~(\ref{cond}).

As before we sum up all the resonant terms
$\sim \cos{\omega t} {\bf u}_{0}({\bf r})$
 and obtain the following equation
for the shift in the oscillation frequency 
\begin{equation}
l(\omega)-l(\omega_{0})=
\frac{a^{2}}{\left ({\bf u}_{0},{\bf u}_{0}\right )}
\left ({\bf u}_{0},
{\bf f}[u_{0}({\bf r})]\right ),
\label{ll}
\end{equation}
where
\begin{eqnarray}
{\bf f}[u_{0}({\bf r})]=
-\frac{1}{2}\frac{\delta {\hat G}_{2}}{\delta {\bf u}}
\bigg |_{{\bf u}={\bf u}_{0}({\bf r})}
\left [
{\hat h}^{-1}+
\frac{1}{2}{\left ({\hat h}-l(2\omega_{0} )\right )}^{-1}\right ]
{\hat G_{2}}\left [{\bf u}_{0}({\bf r})\right ]+
\frac{3}{4}{\hat G}_{3}\
\left [{\bf u}_{0}\left ({\bf r}\right )\right ]
\label{f}
\end{eqnarray}
and $({\bf x},{\bf y})$ denotes a scalar product (see Appendix for a general
discussion).
The criterion for a first-order transition reads
$\tau (a)-\tau (0)<0$ and using ${dl}/{d\omega}<0$ we find
\begin{equation}
\left ({\bf u}_{0}({\bf r}),
{\bf f}[u_{0}({\bf r})]
\right )<0.
\label{crit}
\end{equation}
In the next section we apply the criterion (\ref{crit})
to the tunneling of strings.

\section{tunneling of strings}
\subsection{General theory}

In this section we apply Eqs.~(\ref{f}) and (\ref{crit}) to the problem of
a driven string 
${\it i})$ tunneling between two potential wells and
${\it ii})$
depinning from a columnar defect. Within the linear elasticity
theory the imaginary time Lagrangian of the string takes the form
\begin{equation}
{\cal L}=
\int\limits_{-{L}/{2}}^{{L}/{2}}
dz\left (L_{D}(u,\partial_{t} u)+
\frac{\epsilon}{2}
{\left (
\frac{\partial u}{\partial z}
\right )}^{2}+
V(u)-Fu\right ).
\label{lagr}
\end{equation}
Below we consider the dynamical term
$L_{D}(u,\partial_{t} u)$ to be the sum of massive and dissipative  terms
(we ignore the transverse contribution from the Hall term)
\begin{equation}
L_{D}\left (u,{\partial_{t}}u \right )=
\frac{\rho}{2}{\left ({\partial_{t}}u \right )}^{2}-
\frac{\eta}{2\pi}{\partial_{t}u}
\int\limits_{-{\hbar}/{2 T}}^{+{\hbar}/{2T}}
dt_{1}
\ln \left |\sin \left [\pi\left (t-t_{1}\right )
\frac{T}{\hbar}\right ] \right |{\partial}_{t_{1}}u.
\label{dynamics}
\end{equation}
 $L$ and $\epsilon$ are, respectively,
the length and the elasticity of the string, $V(u)$ is the potential,
the external force is assumed to be small:
$F\ll F_{c}$ with $F_{c}$ the depinning force.
 The thermal saddle-point
solution $u_{th}({\bf r})$ corresponding to the Lagrangian
satisfies the equation
\begin{equation}
\epsilon\frac
{\partial^{2}{u}}{\partial z^{2}}=
\frac{\partial V}{\partial u}-F.
\label{Newt}
\end{equation}

The operators ${\hat h},$ ${\hat G}_{2},$ ${\hat G}_{3},$ and
${\delta {\hat G}_{2}}/{\delta u}|_{u_{0}(z)}$
take the following form,
\begin{equation}
{\hat h}=-\epsilon\frac{\partial^{2}}{\partial z^{2}}+
\frac{\partial^{2}V}{\partial u^{2}}\bigg |_{u_{th}(z)},
\label{sch}
\end{equation}
\begin{equation}
{\hat G}_{2}[y]=\frac{1}{2}\frac{\partial^{3}V}{\partial u^{3}}
\bigg |_{u_{th}}y^{2},\ \
{\hat G}_{3}[y]=\frac{1}{6}\frac{\partial^{4}V}{\partial u^{4}}
\bigg |_{u_{th}}
y^{3},\ \
\frac{\delta {\hat G}_{2}}{\delta u}\bigg |_{u_{0}}(y)=
\frac{\partial^{3}V}{\partial u^{3}}\bigg |_{u_{th}}u_{0}(z)y,
\end{equation}
where $y$ is an arbitrary function of $z.$ The scalar product $(y_{1},y_{2})$
of two arbitrary functions $y_{1}(z)$ and $y_{2}(z)$ is defined as
$\int_{-{L}/{2}}^{+{L}/{2}}y_{1}(z)y_{2}(z)dz.$ 

Eq.~(\ref{crit}) takes the form
\begin{eqnarray}
& \thinspace &
(u_{0} ,f[u_{0}])=
-\frac{1}{4}\int\limits_{-{L}/{2}}^{{L}/{2}}
dz\frac{\partial^{3}V}{\partial u^{3}}\bigg |_{u_{th}}
u_{0}^{2}
\left ({\hat h}^{-1}+\frac{1}{2}
{\left ({\hat h}-l(2\omega_{0})\right )}^{-1} \right )
\frac{\partial^{3}V}{\partial u^{3}}\bigg |_{u_{th}}
u_{0}^{2}
\nonumber\\
& \thinspace & +\frac{1}{8}
\int\limits_{-{L}/{2}}^{{L}/{2}}
dz\frac{\partial^{4}V}{\partial u^{4}}\bigg |_{u_{th}}
u_{0}^{4}<0,
\label{str}
\end{eqnarray}
where $l(w)=-\rho \omega^{2}-\eta |\omega |.$
It is appropriate to mention that the operator ${\hat h}$ has
one zero eigenvalue in the limit $L\rightarrow\infty,$ 
(see \ref{Motion})
such that
taking its inverse ${\hat h}^{-1}$ needs some care:
The eigenfunction corresponding to the zero eigenvalue
is an odd function of $z,$ whereas
$u_{0}^{2}{\partial^{3} V}/{\partial u^{3}}\big |_{u_{th}}$
is an even function
and there is no contribution to the integral in Eq.~(\ref{str}).
 Consequently, the contribution of the eigenfunction
corresponding to the zero eigenvalue is equal to zero.

Let us investigate the sign of Eq.~(\ref{str}) in the two cases mentioned 
above.

\subsection{Motion of a string across a slightly tilted
periodic potential}
\label{Motion}
The tunneling of a string across a titled periodic potential
is a convenient model for the description of the motion of
 dislocations across the Peierls barrier. 
In the case being studied 
 $V(u)$ is the periodic part of the potential (see Eq.~(\ref{lagr})).
Below we denote time by $t$ and the oscillation period by $\tau.$
This problem shows a smooth crossover from
a purely thermal to a thermal-assisted
quantum activation at a temperature $T_{c}=T_{0}$ of order
$\hbar{({F}/{\rho d})}^{{1}/{2}}$
(see Ref.\cite{Ivlev})
with $d$ the period of the potential
$V(u).$ The schematic evolution of the nucleus is shown in
Figs.~5(a)--5(c).
At temperatures smaller than $T_{1}\sim{\hbar F}/{{(\rho V_{0})}^{{1}/{2}}},$
with $V_{0}$ the amplitude of the potenrial, 
the nucleus is of circular shape (see Fig.~5(a))

\centerline{\epsfxsize=4cm \epsfbox{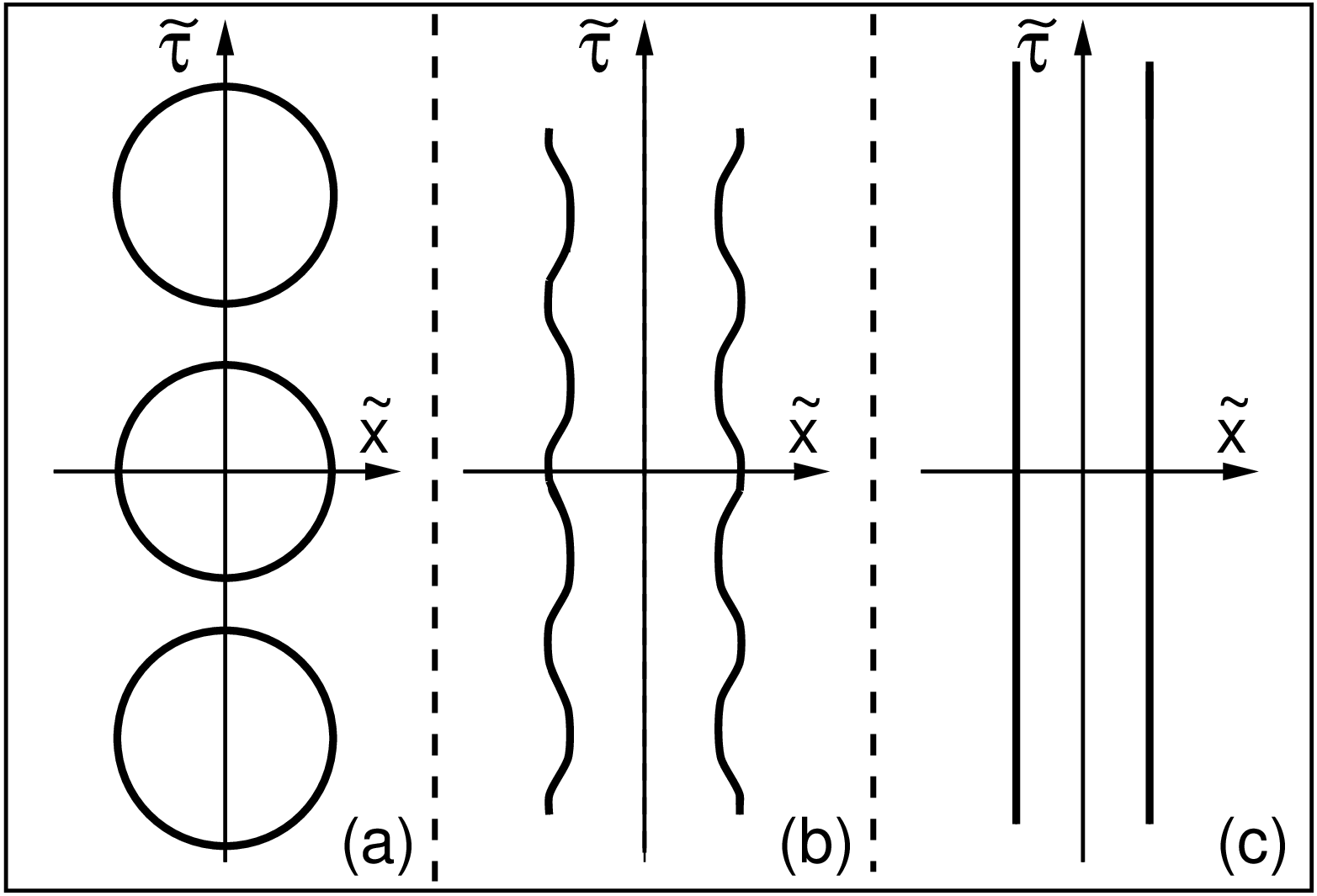}}
{\footnotesize {\bf Fig.5}~
{
 Evolution of the nucleus (in coordinates ${\tilde z}={z}/{\sqrt{\epsilon}}$
and ${\tilde t}={t}/{{\sqrt{\rho}}}$).
(a) Nucleus at low temperatures.
(b) Nucleus at intermediate temperatures where the
interaction between the nucleus' walls is relevant.
(c) Nucleus at high temperatures. The string overcomes the barrier
due to thermal activation.}}
\vskip0.3cm
\hskip-0.7cm
with the bounce solution
$u(z,t)\cong a\Theta(R-r),$ $R\sim{\sqrt{V_{0}}/{F}}$
is the radius of the nucleus ($r^{2}={z^{2}}/{\epsilon}+
{t^{2}}/{\rho}\equiv{\tilde z}^{2}+{\tilde t}^{2}$) and
 the Euclidean action
$S\sim\sqrt{\rho\epsilon}{dV_{0}}/{F}.$
At
$T>T_{1}$ we have to take into account the interaction of
the nucleus' walls (Fig.~5(b)).
At $T>T_{c}=T_{0}$ the string overcomes the barrier due to
thermal activation with $U\sim d\sqrt{\epsilon V_{0}}$ (Fig.~5(c)).
The thermal solution then can be obtained by a continuous
deformation of the zero temperature bounce.
As the problem exhibits a second-order transition from quantum
to classical behavior, we should expect $l(\omega)-l(\omega_{0})>0.$
Let us prove this inequality.
The ``potential'' in the one-dimensional Schr\"odinger operator
${\hat h}$ (see (\ref{sch})) is shown in Fig.~6.
The spectrum of ${\hat h}$ consists of one negative, one zero
(in the limit $L\rightarrow\infty$), and positive eigenvalues
(one can easily verify that the function
${\partial u_{th}}/{\partial z}$
is an eigenfunction of the operator ${\hat h}$ with zero eigenvalue;
on the other hand,
${\partial u_{th}}/{\partial z}$
has one node, i.e., there is one eigenfunction with a negative eigenvalue).
In the limit $F\rightarrow 0$ the negative eigenvalue 
$h_{0}$
tends to zero
as $-F,$ 
see Ref.\cite{Ivlev}. 
The positive eigenvalues of the discrete spectrum of
the operator ${\hat h}$ are of order ${V_{0}}/{d^{2}}.$
Analysing the terms in Eq.~(\ref{str}) we note that the operator
$\left [{\hat h}^{-1}+{1}/{2}{\left ({\hat h}-l(2\omega_{0})\right )}^{-1}
\right ]$
has a large negative eigenvalue $\sim -{1}/{F}$ originating
from the unstable direction in the phase space. The remaining spectrum
is positive and does not depend on $F,$ as is also the 
case for the other factors appearing in Eq.~(\ref{str}).

\centerline{\epsfxsize=7cm \epsfbox{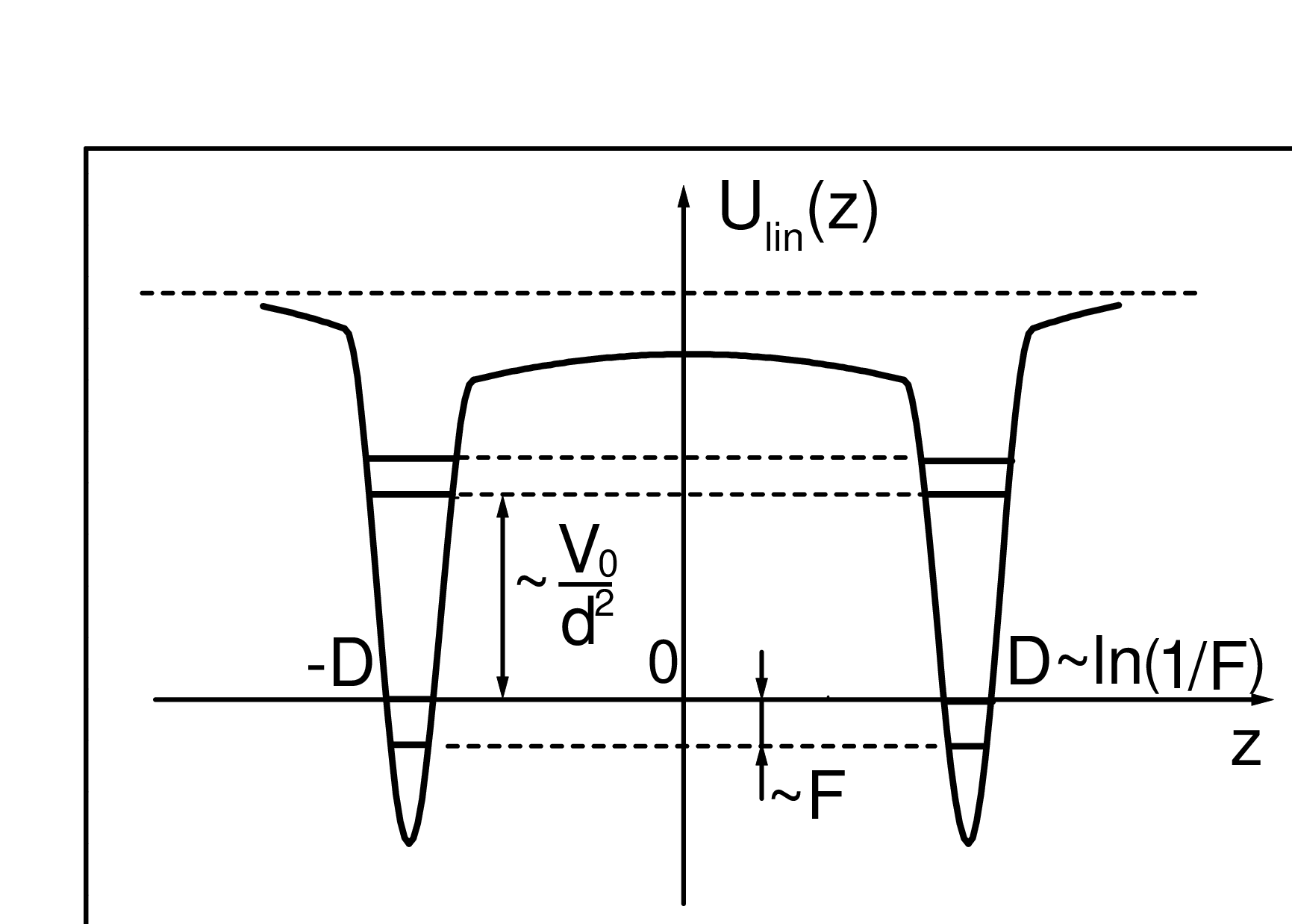}}
\vskip1cm
{\footnotesize {\bf Fig.6}~
{The potential in the 1D Schr\"odinger operator
for the linearized problem of a string tunneling of a string
across a slightly titled periodic potential.
$U_{lin}(z)={\partial^2 U}/{\partial z^2}|_{u_{th}(z)}.$
The negative eigenvalue appears because of the interaction
of the two wells. The distance between the negative
and zero eigenvalues 
is much smaller than the distance between possible positive eigenvalues
of the discrete spectrum and the zero-eigenvalue level.} 
}
\vskip0.3cm
\hskip-0.7cm
Thus we can approximate $\left (u_{0},f[u_{0}]\right )$
by the contribution from the unstable direction and arrive at the estimate
\begin{equation}
(u_{0},f[u_{0}])\simeq
-\frac{1}{4}
\left (
\frac{1}{h_{0}}+
\frac{1}{2\left (h_{0}-l(2\omega_{0})\right )}
\right )
{\left (\int\limits_{-{L}/{2}}^{{L}/{2}}
\frac{\partial^{3}V}{\partial u^{3}}u_{0}^{3}dz\right )}^{2}.
\end{equation}
The second term in Eq.~(\ref{str}) does not involve the small 
parameter $F$ and can always be neglected in the limit $F\rightarrow 0.$
As $l(\omega )=-\rho \omega^{2}-\eta |\omega|,$
$h_{0}=l(\omega_{0})$ we obtain
\begin{equation}
\frac{1}{h_{0}}+
\frac{1}{2\left (h_{0}-l(2\omega_{0}) \right )}<0
\end{equation}
and, consequently, $l(\omega)-l(\omega_{0})>0.$
We have shown therefore that in the vicinity of the point $a=0,$
$\tau (a)$ is an increasing function of $a.$ This is in agreement 
with the fact that the problem of the tunneling of a massive
string across a slightly titled periodic potential exhibits a
second-order transition\cite{Ivlev}.

\subsection{Depinning of a string from a linear defect}
\label{lineardef}

The behavior described above
should be contrasted with the sharp transition
obtained recently for the tunneling of a massive string from a linear
defect \cite{Skvortsov}.
 The imaginary time Lagrangian has the form ($\ref{lagr}$) 
($\eta =0$) with the potential
$U(u)$ describing a single potential well of depth $V_{0}$
and radius $d.$
At low temperatures the nucleus has a circular shape as well.
The
macroscopic part of the
bounce solution can be written in the form
\begin{equation}
u(z,t)=-\frac{F}{4}r^{2}
+\frac{2V_{0}}{F},
\ \ r^{2}=
\frac{z^{2}}{\epsilon}+
\frac{t^{2}}{\rho}\equiv
{\tilde z}^{2}+{\tilde t}^{\thinspace 2}.
\label{bounce}
\end{equation}
The substitution of Eq.~($\ref{bounce}$)
into Eq.~($\ref{lagr}$) gives the Euclidean action
\begin{equation}
\frac{S_{q}}{\hbar}=\frac{1}{\hbar}\int
L[u(z,t)]dt=\frac{4\pi}{\hbar}\sqrt{\rho\epsilon}
{\left (\frac{V_{0}}{F} \right )}^{2}.
\end{equation}
The thermal exponent is given by
${U}/{T}=({4\sqrt{2}}/{3})\sqrt{\epsilon V_{0}}
({V_{0}}/{FT}).$
${S_{q}}/{\hbar}$ and ${U}/{T}$ become equal at
$T_{c}=({\sqrt{2}}/{3\pi}){\hbar F}/{{(\rho V_{0})}}^{{1}/{2}}.$
On the other hand, the dynamical solution ($\ref{bounce}$) remains valid
as long as the radius of the nucleus is smaller than the periodicity in $t,$
i.e., for $T<T_{1}={{3\pi}/{8}}T_{c}.$
 The formal application of the perturbative
procedure (see section~\ref{nonlinear})
 gives the following linearized problem
determining the crossover temperature $T_{0},$
\begin{equation}
-\epsilon
\frac{{\partial}^{2}u_{0}}{\partial z^{2}}+
\frac{{\partial}^{2}V}{\partial u^{2}}\bigg |_{u_{th}(z)}u_{0}=
-\rho{\left (\frac{2\pi T_{0}}{{\hbar}}\right )}^{2}u_{0} .
\label{eigg}
\end{equation}
The 1D Schr\"odinger operator in (\ref{eigg}) has again one negative eigenvalue
which does not depend on the details of the potential
$V(u)$ in the limit $F\rightarrow 0$ (Ref.~\cite{Skvortsov,Kramer}).
$T_{0}$ then is given by the formula
$T_{0}=({\mu}/{2^{{3}/{2}}\pi}){\hbar F}/{\sqrt{\rho V_{0}}}=0.900T_{c}$
with $\mu=1.199$ the root of the equation $\mu\tanh{\mu}=1,$
see Ref.\cite{Kramer}.
As $T_{0}<T_{c}<T_{1},$
the string exhibits a first-order transition at the temperature
$T_{c}.$ In this case the quantum solution (see  Fig.~5(a))
jumps at $T_{c}$ to the thermal solution (Fig.~5(c)).
The intermediate regime (Fig.~5(b)) is absent.

This
sharp transition is again in agreement with the above considerations:
There exist solutions of the equation of motion in imaginary time
with periods
$\tau\in ({{\hbar}/{T_{1}}}, \infty)$
and energy close
to zero (we consider a massive dynamics here).
On the other hand, at the temperature $T_{0}$ there appears a
time-dependent solution
of the equation of motion with a period
${\hbar}/{T_{0}}$
and an energy equal to the energy of the thermal solution $U.$
As ${\hbar}/{T_{1}}<{\hbar}/{T_{0}},$
the dependence of the oscillation period on energy is not monotonous
and a first-order transition takes place.
In the present case the jump of the
derivative
${\left [ E_{1}(T_{c})-U \right ]}/{{T_{c}}^{2}}$ in the action is large:
$E_{1}(T_{c})$ is close to zero, $U$ and $T_{c}^{-1}$
are proportional to ${1}/{F},$ hence
$\Delta ({\partial S_{\rm Eucl}}/{\partial T})\sim -{1}/{F^{3}}$
and one can see that we deal with a strong first-order transition.

Let us find the sign of $l(\omega)-l(\omega_{0})$ in
(\ref{ll})
for the case $\eta\ne 0,$ $\rho\ne 0.$ The ``potential'' in the one-dimensional Schr\"odinger
operator ${\hat h}$ is shown in Fig.~7. 

\centerline{\epsfxsize=7cm \epsfbox{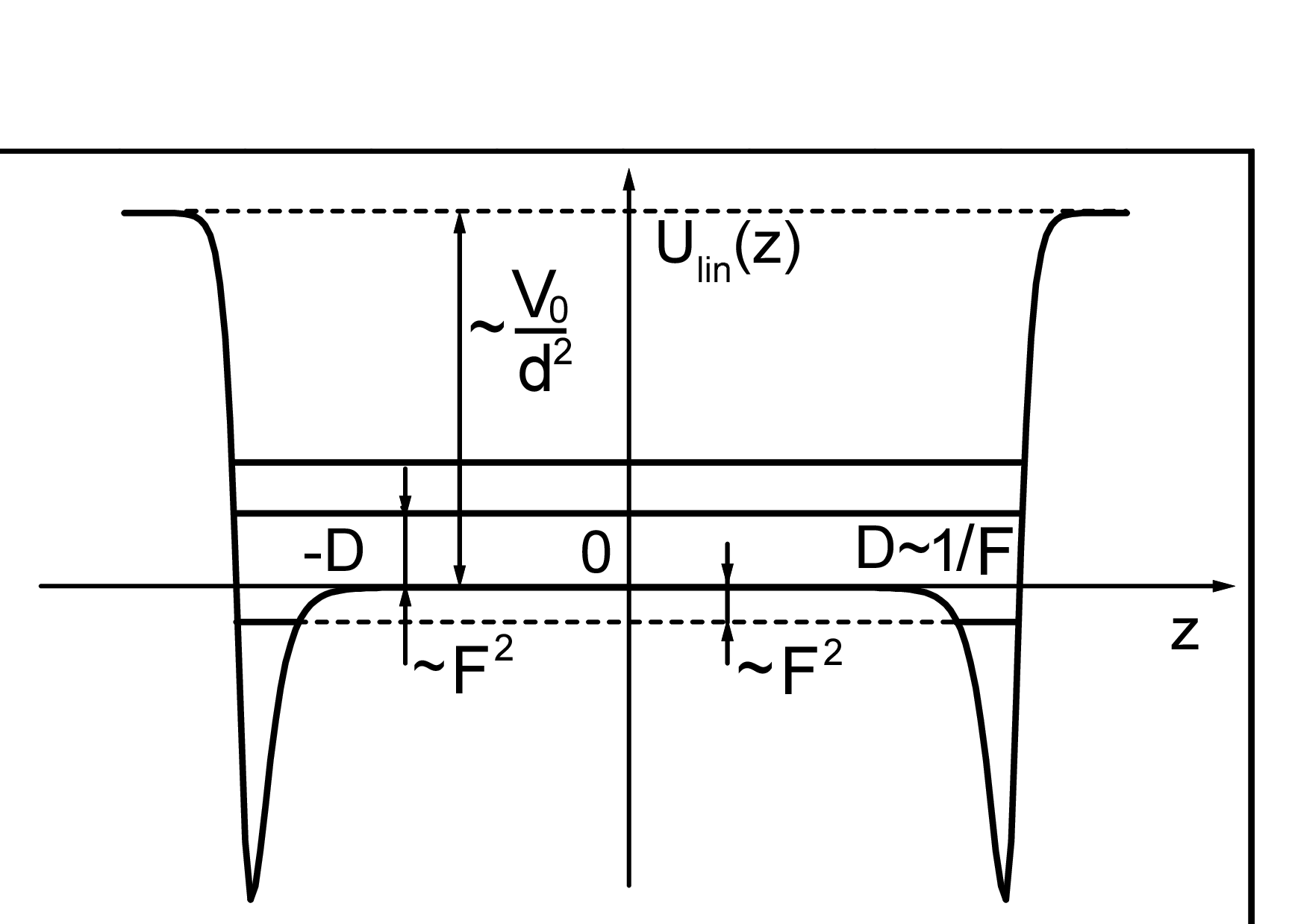}}
\vskip1cm
{\footnotesize {\bf Fig.7}~
{The potential in the 1D Schr\"odinger operator
for the linearized problem of the depinning of a string
from a linear defect.
$U_{lin}(z)={\partial^2 U}/{\partial z^2}|_{u_{th}(z)}.$
The modulus of the negative eigenvalue is of the same 
order as positive eigenvalues in the discrete spectrum.}
}
\vskip0.3cm
\hskip-0.7cm
In the limit $F\rightarrow 0$
the most
 significant contribution to the integrals in (\ref{str})
arises from the region $|z|>D={\sqrt{2\epsilon U_{0}}}/{F}.$
The spectrum of the operator $\hat h$ consists of one negative, one zero, and
positive eigenvalues. In the limit 
$F\rightarrow 0$ the negative eigenvalue and
the lowest positive eigenvalues of the descrete spectrum are proportional to
$F^{2}$ and we need to analyze the first integral in Eq.~(\ref{str})
more carefully than before. The second integral in Eq.~(\ref{str})
is again irrelevant. In the limit $F\rightarrow 0,$ the eigenfunctions
of the lowest levels take the same 
form in the region $|z|>D,$ i.e., we can write
\begin{eqnarray}
(u_{0},f[u_{0}])
& \thinspace & \cong -\frac{1}{4}\int\limits_{-{L}/{2}}^{{L}/{2}}dz
\frac{\partial^{3}V}{\partial u^{3}}\bigg |_{u_{th}}
u_{0}^{2}
\left ({\hat h}^{-1}+\frac{1}{2}
{\left ({\hat h}-l(2\omega_{0})\right )}^{-1} \right )
\frac{\partial^{3}V}{\partial u^{3}}\bigg |_{u_{th}}
u_{0}^{2}
\nonumber\\
& \thinspace &
=-\frac{1}{4}\int\limits_{-{L}/{2}}^{{L}/{2}}
dz
\frac{\partial^{3}V}{\partial u^{3}}\bigg |_{u_{th}}
u_{0}^{2}
\left (
\sum\limits_{n}
\left ({h_{n}}^{-1}+\frac{1}{2}
{\left ({ h_{n}}-l(2\omega_{0})\right )}^{-1}\right ) 
|u_{n}\rangle\langle u_{n} | \right )
\frac{\partial^{3}V}{\partial u^{3}}\bigg |_{u_{th}}
u_{0}^{2}
\nonumber\\
& \thinspace &
\cong -\frac{1}{4}\sum\limits_{n}
\left ({h_{n}}^{-1}+\frac{1}{2}
{\left ({ h_{n}}-l(2\omega_{0})\right )}^{-1} \right )
{\left (\int\limits_{|z|>D}dzu_{n}
\frac{\partial^{3}V}{\partial u^{3}}\bigg |_{u_{th}}
u_{0}^{2}
\right )}^{2}.
\end{eqnarray}
Here, we have used 
that ${\partial^{3}V}/{\partial u^{3}}|_{u_{th}}\simeq 0$ for
$|z|<0.$ In the region $|z|>D,$ $u_{n}=\pm {C_{n}u_{0}}/{C_{0}}$
with $C_{n}$ denoting the normalizers. Consequently, 
\begin{equation}
(u_{0},f[u_{0}])\cong -\frac{1}{4}
{\left (
\int\limits_{|z|>D}
\frac{\partial^{3}V}{\partial u^{3}}\bigg |_{u_{th}}
u_{0}^{3}dz
\right )}^{2}
\sum\limits_{n}
C_{n}^{2}
\left ({h_{n}}^{-1}+\frac{1}{2}
{\left ({\hat h_{n}}-l(2\omega_{0})\right )}^{-1} \right ).
\end{equation}
For the lowest eigenvalue we have 
$u_{0}(|z|<D)=C_{0}\cosh{\left (\sqrt{\frac{|h_{0}|}{\epsilon}z}\right )}$
 and the eigenfunction belonging to
the zero eigenvalue takes the form $u_{1}(|z|<D)=C_{1}z.$ 
 In the limit $F\rightarrow 0$
we have
${u_{n}^{\prime}/{u_{n}}|_{z=D^{-}}}=
{u_{n}^{\prime}/{u_{n}}|_{z=D^{+}}}=
{u_{0}^{\prime}}/{u_{0}}|_{z=D^{+}},$ i.e., ${u_{n}^{\prime}}
/{u_{n}}|_{z=D^{-}}=
{u_{0}^{\prime}}/{u_{0}}|_{z=D^{-}}$
(we use the condition of  continuity of the logarithm derivative),
i.e., for the lowest eigenvalue we obtain the condition
\begin{equation}
\sqrt{\frac{|h_{0}|}{\epsilon}}D
\tanh{\left (
\sqrt{\frac{|h_{0}|}{\epsilon}}D
\right )}=1.
\label{zzero}
\end{equation}

Only even levels contribute to the integral in (\ref{str})
and their eigenfunctions are given by
$C_{n}\cos{\left (\sqrt{\frac{h_{n}}{\epsilon}}z\right )},$
hence we obtain for $h_{n}$
\begin{equation}
\sqrt{\frac{h_{n}}{\epsilon}}D
\tan{\left (
\sqrt{\frac{h_{n}}{\epsilon}}D
\right )}=-1.
\label{Even}
\end{equation}
The most significant contribution to the integrals
$\int_{-{L}/{2}}^{{L}/{2}}u_{n}^{2}dz$ arises from the interval
$|z|<D,$ i.e.,
\begin{equation}
C_{0}^{2}=\frac{1}
{\int\limits_{-D}^{D}
\cosh^{2}
{\left (
\sqrt{\frac{|h_{o}|}{\epsilon}}z
\right )}dz}=
\frac{1}{D\cosh^{2}
{\left (
\sqrt{\frac{|h_{o}|}{\epsilon}}D
\right )}}
\label{0}
\end{equation}
and
\begin{equation}
C_{n}^{2}=\frac{1}
{\int\limits_{-D}^{D}
\cos^{2}
{\left (
\sqrt{\frac{h_{n}}{\epsilon}}z
\right )}dz}=
\frac{1}{D\cos^{2}
{\left (
\sqrt{\frac{h_{n}}{\epsilon}}D
\right )}},\ \ n=2,4\dots
\label{n}
\end{equation}
In  Eqs.~(\ref{0}) and (\ref{n}) 
we have used Eqs.~(\ref{zzero}) and (\ref{Even}).

Let us suppose that $\eta=0.$ Later we will show that if the problem
exhibits a first-order transition for $\eta =0,$ the kind
of transition remains the same for any $\eta\ne 0.$
Numerical solution of Eqs.~(\ref{zzero}) and (\ref{Even})
show that $h_{0}=-1.439229{\epsilon}/{D^{2}}$ and
$h_{2}=7.830964{\epsilon}/{D^{2}}.$ For the case of a
purely massive dynamics $l(2\omega_{0})=4l(\omega_{0})=4h_{0}.$
The operator
${\hat h}^{-1}+({1}/{2}){\left ({\hat h}-l(2\omega_{0})\right )}^{-1}$
has only one negative eigenvalue, i.e., if we show that
the sum
\begin{equation}
S=\sum\limits_{n}^{M}
C_{n}^{2}
\left (
\frac{1}{h_{n}}+
\frac{1}{2\left (h_{n}-l(2\omega_{0})\right )}
\right ),\ \ n=0,2\dots
\end{equation}
is positive for some finite $M,$
a first-order transition takes place. With $M=2$ we obtain
\begin{equation}
C_{0}^{2}
\left (
\frac{1}{h_{0}}+
\frac{1}{2\left (h_{0}-l(2\omega_{0})\right )}
\right )+
C_{2}^{2}
\left (
\frac{1}{h_{2}}+
\frac{1}{2\left (h_{2}-l(2\omega_{0})\right )}
\right )=0.008796{D}/{\epsilon}>0,
\end{equation}
i.e., the transition from quantum to classical behavior
is of first-order.

Finally, let us show that the kind of transition remains the same
for any $\eta\ne 0.$ We can write for the ratio
\begin{equation}
\frac
{l(2\omega_{0})}
{l(\omega_{0})}=
\frac
{4\rho{\omega_{0}}^{2}+
2\eta{\omega_{0}}  }
{\rho{\omega_{0}}^{2}+
\eta{\omega_{0}} }
=4-
\frac{2\eta\omega_{0}}
{\rho{\omega_{0}}^{2}+
\eta{\omega_{0}}}<4,
\end{equation}
i.e., the relative contribution of the operator
$({1}/{2}){\left ({\hat h}-l(2\omega_{0})\right )}^{-1}$ becomes even  larger
compared to the contribution of the operator ${\hat h}^{-1}.$

Comparing the problems considered in sections \ref{Motion}
and \ref{lineardef} the main difference appears because of the different
character of the spectrum of the linearized problem.
For the case of a string tunneling across a slightly titled periodic potential
the most significant contribution arises from the negative eigenvalue.
The contributions of the positive eigenvalues are negligible in the limit
$F\rightarrow 0.$ 
On the other hand, for a string depinning from a linear defect
the contributions of the positive eigenvalues are of the same order as
that of the negative eigenvalue.

\section{conclusion}
\label{conclusion}

If a second order transition takes place, it is possible to use a
perturbative approach for the calculation of the crossover temperature.
In this case we have to substitute equation (\ref{40})
into the equation of motion. Close to $T_{c}$ the function
$\delta {\bf u}(\bf r)$ is small and we obtain
the linearized eigenvalue problem (see also Eq.~(\ref{eig}))
\begin{equation}
\frac{\delta^{2}S_{\rm Eucl}[{\bf u}]}{\delta {\bf u}^{2}}\bigg |_
{{\bf u}={\bf u}_{0}({\bf r})}\delta {\bf u}=\lambda \delta{\bf u},
\label{exp}
\end{equation}
where the only negative eigenvalue of Eq.~(\ref{exp})
(due to the single unstable direction in the phase space)
 determines the transition
temperature
$T_{c}.$ For the case of a first order transition at $T_{c}$ the formal
application of the perturbative procedure again
 gives some ``crossover"
temperature $T_{0}$ which comes to lie below the true transition
at $T_{c},$ however. For the
dependence of the oscillation time on energy
considered above
(see Fig.~4) $T_{0}={\hbar}/{\tau_{0}}.$ In Fig.~8 we plot
the actions for the various extremal solutions corresponding to the
situation in Fig.~4.
 
\centerline{\epsfxsize=4cm \epsfbox{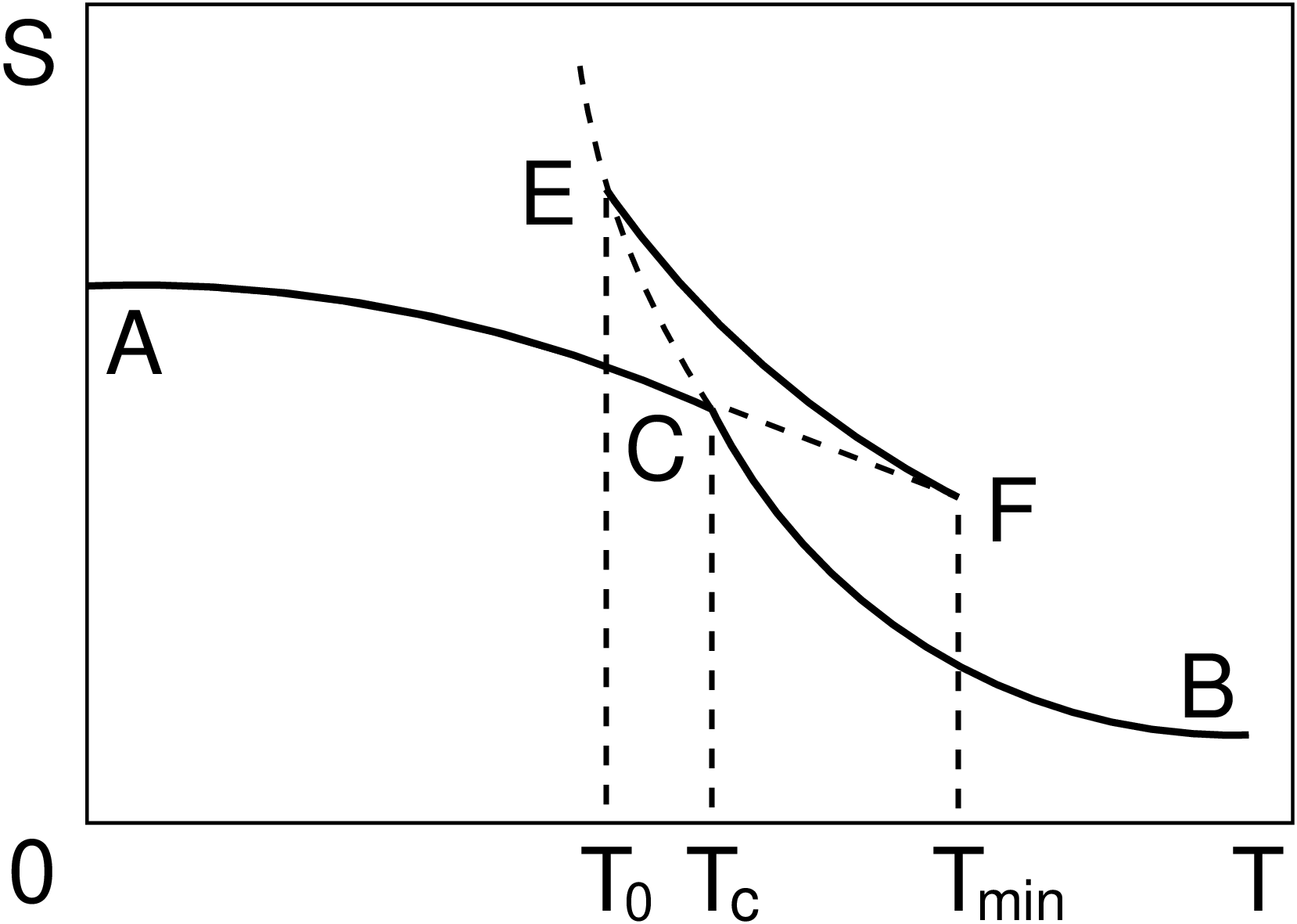}}
{\footnotesize {\bf Fig.8}~
{Euclidean action $S$ as a function of temperature $T$
for a non-monotonous oscillation time
 $\tau (E)$ as given in Fig.~4.}
}
\vskip0.3cm
\hskip-0.7cm
The solid line $AC$ corresponds to quantum tunneling. At the point $C,$
a first-order transition takes place
and the solid line $CB$ shows the thermal action.
In the point $E$ the ``false" crossover takes place.
 The line $EF$ corresponds to the
oscillating motion of the tunneling system with energy
$E\in (E_{\rm min}, U_{0})$ (see Fig.~4).
The dashed line $CF$ is obtained by the continuation of the quantum action into
the region $T>T_{c}$ and corresponds to the motion with energy
$E\in (E_{1}(T_{c}), E_{\rm min}).$
In the point $F$ this line and the ``false"
crossover action intersect each other. The dashed line $EC$ is a formal continuation
of the thermal action into the region $T<T_{c},$ see also 
Ref.\cite{Chudnovsky}.

The above discussion shows that
in the case of a first-order transition one cannot
use the perturbative representation of the bounce solution
(see equation $(\ref{40})$). The transition takes place at a temperature
higher than that given by the formal solution of the
linearized eigenvalue
problem. In this case one has to find the low temperature
expression for the Euclidean action and determine the point where it is equal
to the thermal action.

There are a number of articles
concerning the decay of metastable states, where a sharp
transition from quantum to classical behavior
has been discovered.
A first-order transition has been obtained
by Lifshitz and Kagan in Ref.~\cite{Lifshitz}
concerning phase transitions at low temperatures,
by Chudnovsky\cite{Chudnovsky} for a massive particle in a 1D potential
for the case of a nonmonotonous dependence of the imaginary time
oscillation period on energy,
by Morais-Smith, Ivlev, and Blatter in Ref.~\cite{CMS}
where macroscopic quantum tunneling in a dc SQUID has been
investigated, by Carriga in Ref.\cite{Carriga} concerning
the vacuum decay in 2 and 3 dimensions in the thin wall limit,
by Ferrera\cite{Ferrera} for a bubble nucleation
in the  $\phi^{4}$ model, and 
by Skvortsov~\cite{Skvortsov} for the case of the depinning of a massive
string from a linear defect.

It would be interesting to observe a first-order transition
in an experiment. Measurements of the escape rate of Josephson junctions
from their zero voltage state\cite{Clarke}
show that the transition from quantum to classical behavior
is of the second kind.
The criterion described above allows us to find
metastable systems exhibiting a sharp transition
without a complicated solution of the equation of motion in the 
whole temperature range.
The problem considered in section~\ref{lineardef}
of the depinning of a string with finite mass and friction coefficients
is related to the problem of the depinning of vortices from columnar defects
produced by the irradiation of heavy ions in high-$T_{c}$ superconductors.
As it has been shown, this problem exhibits a first-order transition.
In principle, this sharp behavior can be observed in an experiment.
However, one should take into account that the point of transition
is smeared due to the different radii of the defects.
Hence, samples with defects of one radius should be taken for an experiment.
An appropriate method to produce defects with identical radii
is based on lithographic technique, 
see Ref.\cite{Dolan,Moshchalkov,Metlushko,Baert}.

Finally, let us discuss the $F-T$ ``phase diagram" of a 
massive string
depinning from a columnar defect. At small $F,$ $T_{c}$ increases
$\sim F$ and the transition is first-order like.
At $F\alt F_{c},$ with $F_{c}$ the critical force, the problem
exhibits a second-order transition with
$T_{c}=T_{0}\sim\sqrt{F_{c}-F},$ see Ref.\cite{Chudnovsky1}. 
Consequently, there is a ``tricritical" point 
$C$
(see also Ref.\cite{Skvortsov}),
where the nature of the transition changes, see Fig.~9.

\centerline{\epsfxsize=7cm \epsfbox{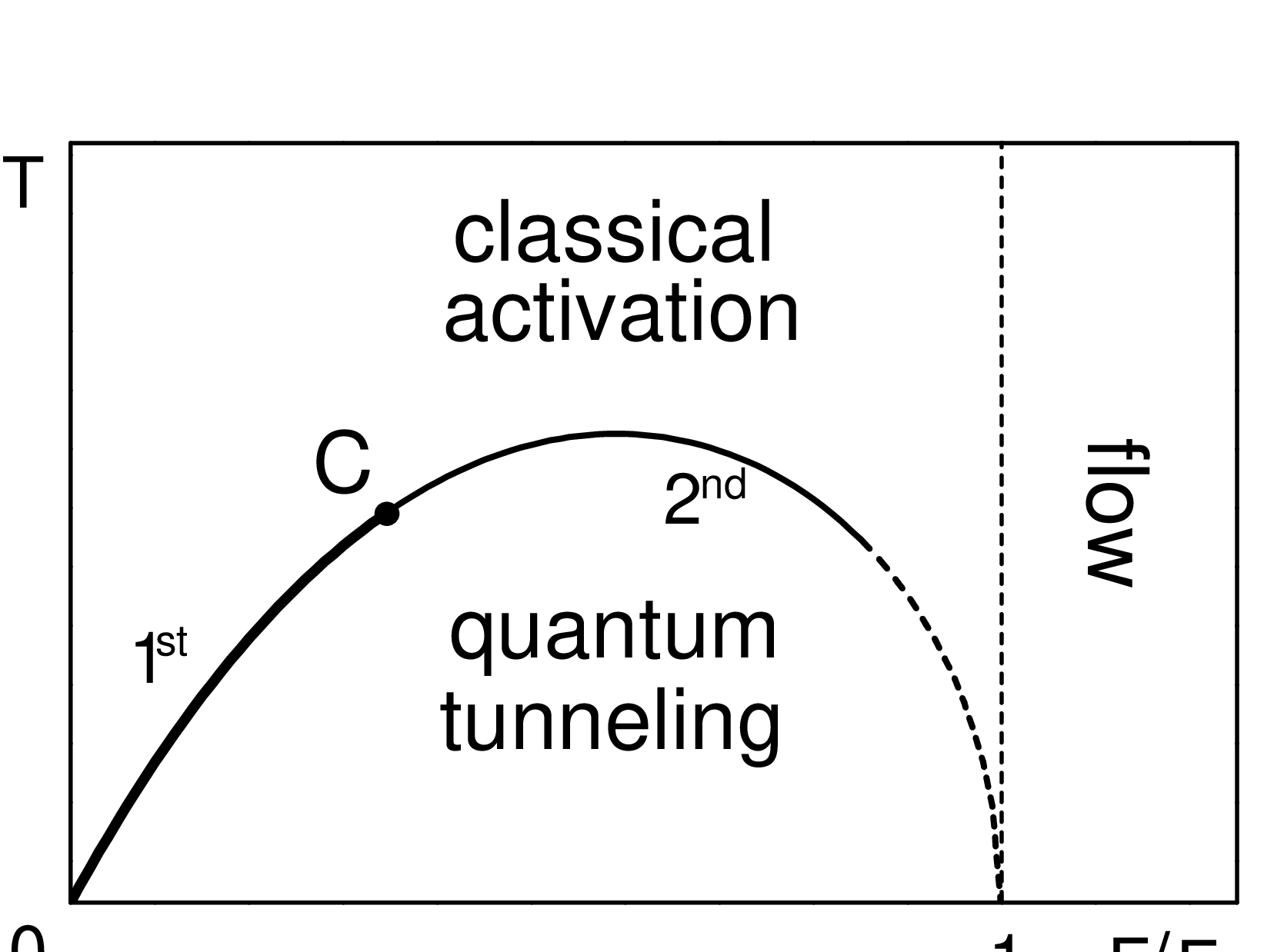}}
\vskip0.5cm
{\footnotesize {\bf Fig.9}~{Possible 
$F$--$T$
``phase-diagram" of a massive string depinning from a
columnar defect. In $C$ the kind of transition changes
from first- to second-order,
i.e., it is a tricritical point. In the vicinity of the point
${F}/{F_{c}}=1$ the curve is dashed as in the limit $F\rightarrow F_{c}$
the barrier disappears and the semiclassical approximation is not applicable.}
}
\vskip0.3cm
\hskip-0.7cm
In the vicinity of the point ${F}/{F_{c}}=1$ the transition curve
is shown to be dashed as in the limit 
$F\rightarrow F_{c}$    
the potential barrier disappears
and the semiclassical approximation is not applicable.

As it has already been mentioned, the kind of transition is determined
by the ``competition" of the negative and positive eigenvalues
of the operator 
$\left ({\hat h}^{-1}+({1}/{2}){({\hat h}-l(2\omega_{0} ) )}^{-1}\right ),$
see section~\ref{Motion}. Let us investigate possible spectra of
this operator for the case of the tunneling of a string.
Consider a 1D Schr\"odinger operator
${\hat h}=-\epsilon{\partial^{2}}/{\partial z^{2}}+U_{lin}(z)$
of the linearized problem, where the function $U_{lin}(z)$
is such that the operator ${\hat h}$ has one negative and one zero eigenvalue.
Let us try to reconstruct the initial metastable potential $V(u).$
As it has been mentioned in section~\ref{Motion},
the function $\Psi ={\partial u_{th}}/{\partial z}$ is
 a solution of the equation ${\hat h}\Psi =0,$ i.e., $u_{th}(z)$
satisfies the equation
\begin{equation}
\frac{\epsilon}{2}
{\left (\frac{\partial u_{th}}{\partial z}\right )}^{2}-
V\left (u_{th}(z)\right )=0,
\end{equation}
where we used the boundary condition $u_{th},u_{th}^{\prime}\rightarrow 0$
if $|z|\rightarrow\infty$ and $V(0),V^{\prime}(0)=0.$ 
Inverting  the equation $u=u_{th}(z)\rightarrow z=u^{-1}_{th}(u)$
we obtain 
\begin{equation}
U(u)=\frac{\epsilon}{2}
{\left ({\frac{\partial u_{th}(z)}{\partial z}}\right )}^{2}
\bigg |_{z=u_{th}^{-1}(u)}.
\end{equation}
Hence, we have shown that for an arbitrary 1D Schr\"odinger
operator of the linearized problem
there is only one corresponding ``original" metastable potential
$V(u).$ We have shown therefore that the mapping $V(u)\rightarrow {\hat h}$
is one-to-one. Consequently, one can construct a potential $V(u)$
such that the linearized ``quantum mechanical" problem has 
the appropriate
spectrum. Let the potential
$V(u)$
 be dependent on some parameter ${\alpha}$
which might be multicomponent, in general.
Changing this parameter we are modifying the spectrum of the operator
$\left ({\hat h}^{-1}+({1}/{2}){({\hat h}-l(2\omega_{0} ) )}^{-1}\right )$
and thus
we can tune the kind of the transition from quantum to classical
behavior. In general the ``phase diagram" might be more complicated
than that in Fig.~9.

Briefly summarizing, we have studied the behavior of the imaginary 
time oscillation period
of a metastable system
 in the vicinity of the saddle-point solution
and have derived a sufficient criterion for a sharp 
first-order transition from quantum to thermal
decay of a metastable state.
The results have been applied to a comparative study of the tunneling
of a massive string across a slightly titled periodic potential
and the depinning of a string from a columnar defect in the 
presence of arbitrarily strong dissipation.
The former problem shows a positive derivative of the oscillation period
with respect to the amplitude of the motion,
in agreement with the second-order
transition for a purely massive problem\cite{Ivlev},
whereas the latter problem shows a sharp behavior for any mass and friction   
coefficient.  

\acknowledgments
We thank M.A. Skvortsov for helpful discussions and for sending us his
results before publication. One of us (D.A.G.) acknowleges
financial support from the Swiss National Foundation.

\begin{appendix}
\section{Condition of disappearence of  resonant terms}
\label{app}

Consider the following equation 
\begin{equation}
{\hat l}{\bf y}={\hat h}{\bf y}+{\bf s}({\bf r},t),
\label{nachalo}
\end{equation}
where the operators ${\hat l}$ and ${\hat h}$ act on time and spatial
variables, respectively, 
${\bf s}({\bf r},t)$ is a given vector.
The spectrum of the operator ${\hat l}$ is negative and the spectrum
of  ${\hat h}$ shall be nonnegative except for one eigenvalue.
The solution of Eq.~(\ref{nachalo})
takes the form
\begin{equation}
{\bf y}={\left({\hat l}-{\hat h}\right )}^{-1}{\bf s}({\bf r},t).
\label{prodol}
\end{equation}
As $[{\hat l},{\hat h}]=0,$
the eigenfunctions of the operator $({\hat l}-{\hat h})$
are products of the eigenfunctions of the operators
$\hat l$ and $\hat h,$ respectively. The operator $({\hat l}-{\hat h})$
might have one zero eigenvalue in which case its inverse 
${({\hat l}-{\hat h})}^{-1}$ is not properly defined.
One then can use Eq.~(\ref{prodol}) only if
the coefficient 
of the zero eigenvalue eigenfunction of 
the operator $({\hat l}-{\hat h})$
 in the expansion of ${\bf s}({\bf r},t)$
over the eigenfunctions of the operator $({\hat l}-{\hat h})$
is equal to zero, i.e., if 
\begin{equation}
({\bf s}, {\bf V}_{0})=0,
\end{equation}
where $({\bf x},{\bf y})$ is a scalar product
and ${\bf V}_{0}$ is the eigenfunction of the operator
$({\hat l}-{\hat h})$ corresponding to the zero eigenvalue.

\end{appendix}


\begin{thebibliography}{99}
\bibitem{Kramers} H.A. Kramers, Physica {\bf 7}, 284 (1940).
\bibitem{Petukhov} B.V. Petukhov and V.L. Pokrovsky,
Zh. Eksp. Teor. Fiz. {\bf 63}, 634 (1972)
[Sov. Phys. JETP {\bf 36}, 336 (1973)].
\bibitem{Ivlev}
B.I. Ivlev and V.I. Mel'nikov, Phys. Rev. B {\bf 36}, 6889 (1987).
\bibitem{Larkin} A.I. Larkin and Yu.N. Ovchinnikov,
 Phys. Rev. B {\bf 28}, 6281 (1983).
\bibitem{Blatter}
for a review see G. Blatter, M.V. Feigel'man, V.B. Geshkenbein,
A.I. Larkin, and V.M. Vinokur, Rev. Mod. Phys. {\bf 66}, 1125 (1994).
\bibitem{Ovchinnikov} A.I. Larkin and Yu.N. Ovchinnikov,
Pis'ma
Zh. Eksp. Teor. Fiz. {\bf 37}, 322 (1983)
[JETP Lett. {\bf 37}, 382 (1983)].
\bibitem{Lifshitz} I.M. Lifshitz and Yu. Kagan,
Zh. Eksp. Teor. Fiz. {\bf 62}, 1 (1972)
[Sov. Phys. JETP {\bf 35}, 206 (1972)].
\bibitem{Chudnovsky}E. M. Chudnovsky, Phys. Rev. A {\bf 46}, 8011 (1992).
\bibitem{Galitskiy}
V.M. Galitsky, B.M. Karnakov, and V.I. Kogan,
{\it Zadachi po kvantowoi mekhanike,} [in Russian
{(\it Problems in Quantum Mechanics)}] (Nauka, Moscow, 1992).
\bibitem{Tunn}A.I. Larkin and Yu.N. Ovchinnikov,
Zh. Eksp. Teor. Fiz. {\bf 86}, 719 (1984)
[Sov. Phys. JETP {\bf 59}, 420 (1984)].
\bibitem{Mel'nikov} V.I. Mel'nikov, Zh. Eksp. Teor. Fiz. {\bf 87},
663 (1984) [Sov. Phys. JETP {\bf 60}, 380 (1984)].
\bibitem{Meshkov} V.I. Mel'nikov and S.V. Meshkov,
J. Chem. Phys. {\bf 85}, 1018 (1986).
\bibitem{Landau} L.D. Landau and E.M. Lifshitz,
{\it Mechanics,}
Course in Theoretical Physics, Vol.~1
(Pergamon, Oxford, 1960).
\bibitem{Caldeira}A. O. Caldeira and A. J. Leggett, Ann. Phys. (N.Y.)
{\bf 149}, 374 (1983).
\bibitem{Skvortsov} M.A. Skvortsov, Phys. Rev. B {\bf 55}, 515 (1997).
\bibitem{Kramer}A. Kr\"amer and M. L. Kuli\a'c,
Phys. Rev. B {\bf 48}, 9673 (1993).
\bibitem{CMS} C. Morais Smith, B. Ivlev, and G. Blatter,
Phys. Rev. B {\bf 49}, 4033 (1994).
\bibitem{Carriga}J. Carriga, Phys. Rev. D {\bf 49}, 5497 (1994).
\bibitem{Ferrera}A. Ferrera, Phys. Rev. D {\bf 52}, 6717 (1995). 
\bibitem{Clarke}J. Clarke, A. N. Cleland, M. H. Devoret,
D. Esteve, and J. M. Martins, Science {\bf 239}, 992 (1988).
\bibitem{Dolan}G.J. Dolan and J.H. Dunsm\"ur,
Physica (Amsterdam) {\bf 152B}, 7 (1988).
\bibitem{Moshchalkov}V.V. Moshchalkov {\it et. al.}, Phys. Scr.
{\bf T55}, 168 (1994).
\bibitem{Metlushko}V.V. Metlushko, 
M. Baert, R. Jonckheere, V.V. Moshchalkov, and Y. Bruynseraede,
Solid State Commun. {\bf 91},
331 (1994).
\bibitem{Baert}M. Baert,
V.V. Metlushko, R. Jonckheere, V.V. Moshchalkov, and Y. Bruynseraede,
 Phys. Rev. Lett. {\bf 74}, 3269 (1995).
\bibitem{Chudnovsky1}E.M. Chudnovsky, A. Ferrera, and A. Vilenkin,
Phys. Rev. B {\bf 51}, 1181 (1995).



\end{thebibliography}
\end{document}